\documentclass[12pt]{article}

\usepackage{graphicx}
\usepackage{a4}

\usepackage{fancybox}
\usepackage{epsfig}
\usepackage{color}

\usepackage{amsfonts}
\usepackage{mathrsfs}
\usepackage{amssymb}
\usepackage{amsmath}

\usepackage{dcolumn}

\usepackage{bm}

\def\pa{\partial}

\def\al{\alpha}
\def\be{\beta}
\def\ga{\gamma}
\def\de{\delta}

\def\th{\theta}

\def\la{\lambda}

\def\Om{\Omega}

\newcommand{\ben}{\begin{equation}}
\newcommand{\een}{\end{equation}}
\newcommand{\bea}{\begin{eqnarray}}
\newcommand{\eea}{\end{eqnarray}}
\newcommand{\ba}{\begin{array}}
\newcommand{\ea}{\end{array}}
\newcommand{\bit}{\begin{itemize}}
\newcommand{\eit}{\end{itemize}}

\newcommand{\vs}[1]{\vspace{#1 mm}}

\newcommand{\dsl}{\pa \kern-0.5em /}

\begin{document}

\topmargin 0pt \oddsidemargin 0mm

\vspace{2mm}

\begin{center}

{\Large Conformal symmetry in force-free electrodynamics}

\vs{10}

 {\large Huiquan Li \footnote{E-mail: hli@qzc.edu.cn} and Jianyong Wang \footnote{E-mail: 32040@qzc.edu.cn}}

\vspace{6mm}

{\em College of Teacher Education, Quzhou University, \\
Quzhou, Zhejiang 324000, China}

\end{center}

\vs{9}

\begin{abstract}

It is shown that conformal symmetry exists in force-free electrodynamics (FFE) in Minkowski spacetime, a foundational framework for describing magnetospheres around astronomical objects. In force-free magnetospheres, charges are constrained to move along magnetic field lines and experience zero Lorentz force, due to the everywhere perpendicular orientation of electric and magnetic fields. However, a general angle-preserving conformal mapping of force-free fields does not necessarily produce another physically admissible force-free configuration when sources are present. In this work, we demonstrate that such invariance can nevertheless arise for certain choices of the free functions. Specifically, the governing stream equation is shown to be invariant under Möbius transformations. This symmetry reveals a structural linkage between known solutions and, notably, maps the region inside a magnetospheric horizon (the lightsurface) of one solution to the exterior of its dual counterpart, and vice versa.

\end{abstract}



\section{Introduction}
\label{sec:introduction}

Force-free electrodynamics (FFE) describes a regime in which electromagnetic fields dominate the inertia and dynamics of a plasma, such that the Lorentz force on the charged particles vanishes. This approximation holds when the electromagnetic energy density vastly exceeds the plasma rest-mass energy density, rendering particle inertia negligible. Consequently, the system is governed by Maxwell's equations supplemented by the force-free condition, $\textbf{f}=\rho_e\textbf{E}+\textbf{j}\times\textbf{B}=0$, where the current density $\textbf{j}$ is constrained to flow along magnetic surfaces. The relativistic FFE \cite{Goldreich:1969sb} in Minkowski spacetime provides an effective framework for modeling magnetically dominated plasmas in pulsar magnetospheres and relativistic jets.

In stationary and axi-symmetric settings, FFE is governed by a single stream equation. This equation can equivalently be obtained from a two-dimensional worldsheet theory that describes magnetospheric configurations of open and closed magnetic field lines, analogous to a string theory with one spacetime dimension. The stream equation is highly nonlinear and contains two arbitrary functions \cite{Scharlemann1973Aligned,1973Rotating}, making it very challenging to obtain meaningful solutions, whether analytically or numerically. To date, only a handful of exact solutions are known \cite{1973Rotating,Comp_re_2016,2016arXiv160807998P}.

From the point of view of Einstein's equivalence principle between acceleration and gravity, rotationally accelerating magnetic fields are expected to behave like a gravitational system. It has long been recognized that understanding rotating electrodynamics requires an approach based on general relativity \cite{Schiff1939A,1973Which,1980Determination}, rather than a simple frame transformation. The fact that the originally linear Maxwell equations reduce, under the force-free condition, to a unique nonlinear equation indicates that rotating FFE shares properties with gravitational systems, which are typically nonlinear as well. Indeed, when the force-free magnetosphere ceases to rotate, the stream equation becomes linear again. The FFE system also admits geometrical description \cite{Gralla:2014yja,Comp_re_2016}. The four dimensional spacetime can be foliated by two-dimensional field sheets formed by axi-symmetric magnetic field lines.
In rotating FFE, a horizon-like surface, i.e., the lightsurface (LS), emerges, beyond which magnetic field lines rotate faster than the speed of light, making it causally inaccessible to charged particles or waves. By analogy with gravitational systems, we propose that Hawking-like radiation of charged particles should occur from such a lightsurface in rotating magnetic fields \cite{Li:2021aib}. Furthermore, we find that negative energy can arise in a rotating magnetosphere \cite{2021PhRvD.104b3009L}, analogous to the ergosphere of a Kerr black hole, which suggests the possible presence of superradiant effects. Thus, rotating FFE provides another valuable example in which gravitational features can be explored through a more familiar theory, i.e., electrodynamics.

In FFE, charged particles are constrained to move along magnetic field lines and, by definition, experience a vanishing Lorentz force. Consequently, electric and magnetic fields remain everywhere perpendicular in any FFE configuration. The inertia of the charges is neglected, allowing them to be treated as effectively massless particles. It is therefore natural to examine this effectively two-dimensional FFE theory using conformal techniques, which are widely applied in fields such as fluid dynamics, condensed matter physics, and string theory. While it is well established that source-free electrodynamics possesses conformal symmetry, whether such symmetry persists in the nonlinear FFE is unknown. In this work, we demonstrate that a conformal symmetry can indeed arise in Minkowski spacetime for a specific choice of the free functions within the FFE framework.

We first represent the FFE in two-dimensioanl complex coordinates in Section \ref{sec:comFFE} and then discuss the conformal symmetry, relating existing analytical solutions in Section \ref{sec:contrans}. Finally, comments are given in Section \ref{sec:discussion}.

\section{Force-free electrodynamics in complex coordinates}
\label{sec:comFFE}

We consider a stationary and axi-symmetric FFE in the cylinder coordinates $ds^2=dx^2+dy^2+x^2d\varphi^2$. The rotation of the magnetic fields induces perpendicular electric fields with the relation: $dA_0=-\Om(A_\varphi)dA_\varphi$, where $\Om$ is the angular velocity of the magnetic field lines and is a free function of $A_\varphi$. The current $I$ flowing along the poloidal magnetic field lines is also a function of $A_\varphi$. For convenience, we define $\psi\equiv 2\pi A_\varphi$ and $I(\psi)\equiv-2\pi xB_\varphi=-2\pi x(\partial_xA_y-\partial_yA_x)$.

In the unit vector basis of the cylinder coordinates, the EM fields are given by:
\begin{equation}\label{e:}
 \textbf{E}=\frac{1}{2\pi}\Omega(\psi)\left(\partial_x\psi, \partial_y\psi, 0\right),
\end{equation}
\begin{equation}\label{e:}
 \textbf{B}=\frac{1}{2\pi x}\left(-\partial_y\psi, \partial_x\psi, -I(\psi)\right).
\end{equation}
From Maxwell's equations, the charge and current densities in Gaussian units are respectively:
\begin{equation}\label{e:}
 \rho_e
=\frac{1}{8\pi^2}\nabla\cdot\left(\Om\nabla\psi\right),
\end{equation}
\begin{equation}\label{e:}
 j^x=\frac{1}{4\pi}I'B_x, \textrm{ }\textrm{ }\textrm{ } j^y=\frac{1}{4\pi}I'B_y,
\end{equation}
\begin{equation}\label{e:}
 j^\varphi=\frac{1}{8\pi^2}\nabla\cdot\left(\frac{1}{x^2}\nabla\psi\right),
\end{equation}
where $\nabla$ is associated with the two poloidal coordinates $(x,y)$ and the primes stand for the derivative with respect to $\psi$.

For the fields and densities, the Lorentz force in the toroidal direction automatically vanishes: $f^\varphi=0$. The vanishing of the poloidal components $f^x$ and $f^y$ results in the equation
\begin{equation}\label{e:}
 (1-\Omega^2 x^2)\nabla^2\psi-\frac{1+\Omega^2x^2}{x}\partial_x\psi
-\Omega\Omega'x^2\nabla\psi\cdot\nabla\psi=F(\psi),
\end{equation}
where
\begin{equation}\label{e:}
 F(\psi)=-II'.
\end{equation}
This equation is known as the stream equation (or pulsar equation in astrophysics). The surface satisfying $x\Om=1$ is the so-called lightsurface (LS). Outside the LS, the rotation velocity of the magnetic field lines exceeds the speed of light. The LS is a horizon for charged particles and waves that are restricted to move along magnetic field lines. For those particles that cross the LS and reach outside, they will never come back across the LS.
The stream equation can be equivalently derived from the following action
\begin{equation}\label{e:}
 S=-\int\frac{1}{x} \left[ \bigl(1-\Omega(\psi)^2 x^2\bigr)
\nabla\psi\cdot\nabla\psi-I(\psi)^2\right]dxdy.
\end{equation}

We now turn to the complex coordinates $z=x+iy$. In the complex coordinates, the poloidal quantities should be redefined accordingly. The poloidal force-free fields are expressed as:
\begin{equation}
 E_P(z,\bar{z})=E_x+iE_y=\frac{1}{\pi}\Omega\bar{\partial}\psi,
\end{equation}
\begin{equation}
 B_P(z,\bar{z})=B_x+iB_y=\frac{2i}{\pi(z+\bar{z})}\bar{\partial}\psi,
\end{equation}
where $\partial=\partial_z$ and $\bar{\partial}=\partial_{\bar{z}}$.
The charge and current densities are expressed as
\begin{equation}\label{e:chargeden}
 \rho_e=\frac{1}{4\pi^2(z+\bar{z})}\Bigl[\partial\bigl((z+\bar{z})\Omega\bar{\partial}\psi\bigr)+\bar{\partial}\bigl((z+\bar{z})\Omega\partial\psi\bigr)\Bigr],
\end{equation}
\begin{equation}
 j_P(z,\bar{z})=j^x+ij^y=\frac{1}{4\pi}I'B_P(z,\bar{z}),
\end{equation}
\begin{equation}\label{e:polcurden}
 j^\varphi=\frac{1}{\pi^2(z+\bar{z})}\Bigl[\partial\bigl(\frac{1}{z+\bar{z}}\bar{\partial}\psi\bigr)+\bar{\partial}\bigl(\frac{1}{z+\bar{z}}\partial\psi\bigr)\Bigr].
\end{equation}
The densities $\rho_e$ and $j^\varphi$ are self-conjugate and so real.

The stream equation becomes
\begin{equation}\label{e:comstreameq}
 \left[4-\Omega^2(z+\bar{z})^{2}\right]\partial\bar{\partial}\psi-\frac{4+\Omega^2(z+\bar{z})^{2}}{2(z+\bar{z})}(\partial\psi+\bar{\partial}\psi)-\Omega\Omega'(z+\bar{z})^{2}
\partial\psi\bar{\partial}\psi=F(\psi).
\end{equation}
The equation is now symmetrical on the coordinates, invariant under the interchange $z\leftrightarrow\bar{z}$. The worldsheet action takes the form
\begin{equation}\label{e:}
 S=-\int \frac{i}{z+\bar{z}}\Bigl[\bigl(4-\Omega(\psi)^2(z+\bar{z})^2\bigr)\partial\psi\bar{\partial}\psi-I(\psi)^2\Bigr]dzd\bar{z}.
\end{equation}
The conserved energy-momentum tensor from the action is given by:
\begin{equation}
\begin{aligned}
T_{zz} & = -i\frac{4-\Omega^2(z+\bar{z})^2}{2(z+\bar{z})}(\partial\psi)^2,  \\
T_{\bar{z}\bar{z}} &= -i\frac{4-\Omega^2(z+\bar{z})^2}{2(z+\bar{z})}(\bar{\partial}\psi)^2, \\
T_{z\bar{z}}&=-i\frac{I^2}{2(z+\bar{z})},
\end{aligned}
\end{equation}
in comparison to the one in the four dimensional sense.

\section{Conformal transformation}
\label{sec:contrans}

Let us now study the variation of the FFE under a conformal transformation:
\begin{equation}\label{e:}
 z\rightarrow w=f(z),
\end{equation}
where $f(z)$ is an analytical function.

The angles between field lines are preserved by conformal transformations. Let us start with the general stationary and axi-symmetric electrodynamics, in which the angle between electric and magnetic field lines is given by
\begin{equation}
 \cos\de=\frac{\textbf{E}\cdot\textbf{B}}{|\textbf{E}||\textbf{B}|}=-i\frac{\bar{\partial} A_0\partial A_\varphi-\partial A_0\bar{\partial} A_\varphi}{|\partial A_0\partial A_\varphi|}.
\end{equation}
Obviously, it is generally invariant and so the force-free fields are always perpendicular under a conformal transformation. Similarly, it can be verified that the toroidal component of the Lorentz force $f^\varphi$ also remains vanishing under an analytical conformal transformation.

The only non-trivial case is the poloidal component of the Lorentz force:
\begin{equation}\label{e:polforce}
 f_P(z,\bar{z})=f^x+if^y
=\frac{i}{2}(z+\bar{z})\left(j^\varphi-\rho_e\Om-\frac{1}{2\pi^2(z+\bar{z})^2}F\right)B_P=0.
\end{equation}
The vanishing of the part inside the large brace exactly results in the stream equation (\ref{e:comstreameq}). It contains three sources: the corotating charge density, the toroidal and the poloidal current densities, which are usually not conformal invariant. From the expression of the vanishing Lorentz force or the stream equation, it can be seen that this condition requires the stream function to transform via a conformal factor: $\psi\rightarrow g\psi$. In addition, the transformation must leave the $x$ direction invariant, satisfying $z+\bar{z}\propto w+\bar{w}$.

In what follows, we show that the conditions for the conformal invariance of the stream equation are saturated by choosing the free functions to be
\begin{equation}\label{e:choicefun}
 \Om=\frac{a}{\psi^{2}}, \textrm{ }\textrm{ }\textrm{ } F=\frac{b^2}{\psi^3},
\end{equation}
where $a$ and $b$ are arbitrary constants. We consider first the inversion mapping and then the more general M\"{o}bius transformation.

\subsection{Inversion}

It can be proved that the stream equation (\ref{e:comstreameq}) with the functions (\ref{e:choicefun}) is conformal invariant, only multiplied by a total factor $|w|^{3}$, under the following transformation with weight $(-\frac{1}{4},-\frac{1}{4})$:
\begin{equation}\label{e:inversion}
 \psi(z,\bar{z})\rightarrow |z|\psi\left(\frac{1}{z},\frac{1}{\bar{z}}\right)=\frac{1}{|w|}\psi\left(w,\bar{w}\right)\equiv\phi(w,\bar{w}).
\end{equation}
This means that, if $\psi$ is a solution, so is $\phi$. It is easy to see that this transformation is an involution, i.e., performing it twice leads to the original result.

 Under the mapping, the $x$ direction is preserved: $z+\bar{z}\rightarrow (w+\bar{w})/|w|^2$. The charge density (\ref{e:chargeden}) and the toroidal current density (\ref{e:polcurden}) transform respectively as
\begin{equation}\label{e:}
 \rho_e\rightarrow|w|^5\rho_e, \textrm{ } \textrm{ }  \textrm{ }  j^\varphi\rightarrow|w|^7j^\varphi.
\end{equation}
But the local charge $dQ=\rho_edV$ remains invariant since the volume $dV\rightarrow dV/|w|^5$.
Certainly, with the transformations, the Lorentz force (\ref{e:polforce}) remains vanishing and $T_{z\bar{z}}$ remains invariant.

\subsubsection{Non-rotating magnetospheres}

In the non-rotating case with $a=b=0$, the stream equation is linear. The equation is much easier to solve and there are infinitely many solutions. By using the above conformal mapping (\ref{e:inversion}), we can find relations between these solutions. Expressed in spherical coordinates $x=r\sin\th$ and $y=r\cos\th$, the solutions are related via
\begin{equation}\label{e:}
 \psi^{(0)}=1-\cos\th\longleftrightarrow\phi^{(0)}=r(1-\cos\th),
\end{equation}
\begin{equation}\label{e:}
 \psi^{(n)}=r^{-n}\sin\th P_n^1(\cos\th) \longleftrightarrow \phi^{(n)}=r^{n+1}\sin\th P_n^1(\cos\th).  \textrm{ } \textrm{ }  \textrm{ }  (n\geq1)
\end{equation}
Since the equation is linear, any combination of the above solutions is a solution. The general solution can be expressed as:
\begin{equation}\label{e:}
 \psi=\sum_{n=0}^\infty A_n\psi^{(n)}+B_n\phi^{(n)}.
\end{equation}
which is self-dual with $A_n=B_n$ for all $n$. Of course, further conformal transformations of these arbitrary solutions also give rise to solutions.

Another self-dual solution is:
\begin{equation}\label{e:}
 \psi=|z-i|-|z+i|=\sqrt{x^2+(y-1)^2}-\sqrt{x^2+(y+1)^2},
\end{equation}
whose field lines are hyperbolic, displayed in \cite{Comp_re_2016}.

Note that the non-rotating case with $\Om=I=0$ does not mean the theory is source-free. The charge density is generally not zero. For example, for the above hyperbolic solution, the charge density $\rho_e=(1/2\pi^2)(1/|z-i|-1/|z+i|)$. The relevant source in Lorentz force (\ref{e:polforce}) is the rotating charge $\rho_e\Om$, but not $\rho_e$.

\subsubsection{Rotating magnetospheres}

The solutions are rare for the rotating case. Fortunately, there exist dual solutions for the particular choice of the functions (\ref{e:choicefun}) with $b=2a$. The dual solutions, i.e., the $n=1$ case above, in spherical coordinates are:
\begin{equation}\label{e:}
 \psi=x^2=r^2\sin^2\th \longleftrightarrow \phi=\frac{\sin^2\th}{r}.
\end{equation}
The former solution depicts just parallel vertical field lines, while the latter is a standard dipole solution, found in \cite{Comp_re_2016,2016arXiv160807998P}.
The transformation (\ref{e:inversion}) maps the interior of $\psi$ to the exterior of $\phi$, and vice versa. Their LS radii are
\begin{equation}\label{e:}
 r_{\textrm{LS}}(\psi)=\frac{|a|^{\frac{1}{3}}}{\sin\th} \longleftrightarrow r_{\textrm{LS}}(\phi)=\frac{\sin\th}{|a|^{\frac{1}{3}}}.
\end{equation}
So an open LS is mapped to a closed LS.

As expected, the charge and toroidal current densities are rescaled appropriately to guarantee the Lorentz force to vanish:
\begin{equation}\label{e:}
 \rho_e(\psi)=\frac{a}{\pi x^4} \longleftrightarrow \rho_e(\phi)=\widetilde{r}^3\frac{a}{\pi \widetilde{x}^4}=r^{5}\frac{a}{\pi x^4},
\end{equation}
\begin{equation}\label{e:}
 j^\varphi(\psi)=j^\varphi(\phi)=0.
\end{equation}
Here, we discriminate the real coordinates of $z$ and $w$ with the tilded and untilded coordinates, respectively.

\subsection{M\"{o}bius transformation}

Actually, the stream equation also admits the trivial symmetries: $z\rightarrow\la z$ and $z\rightarrow z+i\ga$ \cite{2021PhRvD.104b3009L} for arbitrary $\Omega$ and $I$. Hence, combined with the inversion symmetry, the stream equation should bear the more general M\"{o}bius conformal symmetry. Indeed, the general transformation that preserves the $x$ direction takes the form:
\begin{equation}\label{e:Mobiustrans}
 w=\frac{i\al z+\be}{z+i\gamma},  \textrm{ }\textrm{ }\textrm{ } (\be\neq-\al\ga)
\end{equation}
where $\al$, $\be$ and $\ga$ are real numbers. It can be proved that the equation is conformally invariant under the following transformation:
\begin{equation}\label{e:}
 \psi(z,\bar{z})\rightarrow\phi(w,\bar{w})=\frac{\sqrt{|\be+\al\ga|}}{|w-i\al|}\psi(w,\bar{w}).
\end{equation}

This forms a subgroup of the M\"{o}bius transformation group $PSL(2,{\mathbb{C}})$. Performing the transformation repeatedly leads the vertical solution to an infinite series of de-centered dipole solutions. The first two transformations lead to
\begin{equation}
\begin{aligned}
 r^2\sin^2\th &\rightarrow r^2\sin^2\th\left(\frac{|\be_1+\al_1\ga_1|}{r^2-2r\al_1\cos\th+\al_1^2}\right)^{\frac{3}{2}} \\
&\rightarrow  r^2\sin^2\th\frac{|(\be_1+\al_1\ga_1)(\be_2+\al_2\ga_2)|^{\frac{3}{2}}}{\left[(\al_1+\ga_2)^2r^2+2(\be_2-\al_1\al_2)(\al_1+\ga_2)r\cos\th+(\be_2-\al_1\al_2)^2\right]^{\frac{3}{2}}}.
\end{aligned}
\end{equation}
If the transformation group is $SL(2,{\mathbb{C}})$, i.e., $\be_i+\al_i\ga_i=-1$, the second solution above simplifies to
\begin{equation}\label{e:}
 \frac{x^2}{|\al_1+\ga_2|^3}\left[x^2+\left(y+\frac{\be_2-\al_1\al_2}{\al_1+\ga_2}\right)^2\right]^{-\frac{3}{2}}.
\end{equation}
So it is clearly seen that the solution is just a shift of the dipole center along the rotation axis. It is approximately vertical near the origin and the unique dipole asymptotically.

\section{Discussion}
\label{sec:discussion}

We have shown that the stream equation of FFE in Minkowski spacetime admits the M\"{o}bius conformal symmetry of the form (\ref{e:Mobiustrans}). This symmetry preserves the conformal flatness of the spatial three-metric: $ds^2=dz d\bar{z}+(z+\bar{z})^2d\varphi^2/4$, though the metric itself actually enjoys a larger M\"{o}bius conformal group.
This finding is reminiscent of the conformal symmetries previously found for FFE on near-horizon warped AdS spaces of Kerr black holes \cite{Wang:2014vza, Li:2014bta, Lupsasca:2014pfa}. While such symmetries are expected in source-free electrodynamics, the situation here is different: the FFE system studied is not source-free, even in the non-rotating case, and the stream equation is highly non-linear, containing multiple source contributions.

This symmetry relates the analytical solution found previously. An interesting feature of the inversion symmetry is that it maps the region interior to the LS of one solution to the exterior of its dual solution, and vice versa. The LS acts as a causal horizon for charged particles and waves, analogous to horizons in gravitational systems. The physics inside a horizon typically remains inaccessible to external observers. Such an inside-outside mapping (excluding singular points) could therefore offer insights into physics beyond a horizon.

Conformal techniques may be useful for exploring new solutions in FFE. Since the perpendicularity of electric and magnetic fields is preserved under arbitrary conformal transformations, one can in principle generate different stationary magnetospheric configurations via such mappings. However, perpendicular field lines alone are not sufficient to define a valid FFE configuration. It remains necessary to determine appropriate charge densities and free functions that consistently ensure the vanishing of the poloidal Lorentz force.

\section*{Acknowledgements\markboth{Acknowledgements}{Acknowledgements}}

This work is financially supported by the funding KYQD001224007 (QZU).

\bibliographystyle{JHEP}
\bibliography{b}

\end{document}